\documentstyle[preprint,floats,tighten,prl,aps]{revtex}
\def \pom {{\scriptscriptstyle \kern -0.1em I \kern -0.25em P}}
\input epsf.tex
\def\desepsf(#1 width #2){\epsfxsize=#2 \epsfbox{#1}}
\begin{document}
\preprint{\vbox{
\hbox{ETH-TH/99-06} 
\hbox{February 1999} }}
\draft
\def\as{\alpha_s}
\def\ee{e^+e^-}
\def\qq{q \bar{q}}
\def\lmsb{\Lambda_{\overline{\rm MS}}}
\def\to{\rightarrow}
\def\fb{~{\rm fb}}
\def\pb{~{\rm pb}}
\def\ev{\,{\rm eV}}
\def\kev{\,{\rm KeV}}
\def\mev{\,{\rm MeV}}
\def\gev{\,{\rm GeV}}
\def\GeV{\,{\rm GeV}}
\def\tev{\,{\rm TeV}}

\title{Higgs production with large transverse momentum\\ in hadronic collisions at next-to-leading order
 \footnote{Work partly supported by the EU Fourth Framework Programme `Training and Mobility of Researchers', Network `Quantum Chromodynamics and the Deep Structure of Elementary Particles', contract FMRX-CT98-0194 (DG 12 - MIHT) and the Swiss National Foundation.}}
\author{D.\ de Florian, M.\ Grazzini and Z.\ Kunszt}
\address{Institute of Theoretical Physics, 
ETH, CH-8093 Z\"urich, Switzerland}

\maketitle

\hspace{1cm}

\begin{abstract}
Inclusive associated production  of a light Higgs boson
($m_{\rm H}\le m_t$)  with one jet in $pp$ collisions
is studied in   next-to-leading order QCD.
Transverse momentum  ($p_{\rm T}\geq 30\gev$)
and rapidity distributions of the Higgs boson are calculated 
for the LHC in the large top-quark mass limit.
It is pointed out that, as much as in the case of inclusive
Higgs production, the $K$-factor of this process is 
large ($\approx 1.6$) and depends  weakly 
on the kinematics  in 
a wide range of transverse momentum and rapidity
intervals. 
Our result confirms  previous suggestions 
that the production channel $p+p\to H+{\rm jet}$
 $\to \gamma+\gamma\, +$ jet
gives a measurable signal  for Higgs production at the LHC in the mass
range $100-140\gev$, 
crucial also for the ultimate test of the Minimal Supersymmetric
Standard Model.

\end{abstract} 
\vspace{1cm}

\narrowtext

Recent results  from LEP and the SLC
indicate  that the Higgs boson of the
Standard Model might be light.
A fit to the precision data has given the
values $m_{\rm H}=76^{\small+85}_{\small-47} \gev$, 
corresponding to $m_{\rm H}\le 262\gev$ at the $95\%$ confidence
level, whereas a direct search at LEP200 gives the lower limit
 as  $90\gev\le m_{\rm H}$ \cite{lepsld}. 
In addition, a crucial  theoretical upper limit exists
on the mass of the   light neutral scalar Higgs boson
of the Minimal Supersymmetric Standard Model 
$m_h\le 130\GeV$. It is, 
therefore,  significant that
one attempts to get  the best possible
signals  in the light mass range of $100\gev\le m_{\rm H} \le 140\gev$ at the LHC.
Simulation 
studies carried out by ATLAS and CMS  have shown that assuming a low
integrated luminosity
of $3\cdot 10^4~{\rm pb}^{-1}$,  
 even in the case of  
 the ``gold-plated''    decay channel 
 into two photons, the signal significance ($S/\sqrt{B}$) is
 only around 5 \cite{lhc}. 
This conclusion depends on the value of the 
$K$-factor (a conservative value $K$=1.5 was used)
and some  plausible  assumptions on the  size of the
 background.
The calculation of the next-to-leading order (NLO) 
corrections to  the background
is  not yet complete \cite{aurenche} and the contribution of the
  NNLO subprocess $gg\to\gamma\gamma$ is large \cite{back}.
The complete NNLO analysis
is extremely laborious, but appears to be feasible.
Actually, for the full numerical control
the background has to be calculated in  NNNLO, 
which is completely beyond the scope of presently available
techniques. Fortunately, 
the ambiguity in the value of the background 
 to the signal is suppressed by the square root appearing
in the definition of the signal significance.
This situation, which is not completely
satisfactory, can be improved by
studying  the $\gamma+\gamma{\rm \ + \ jet(s)}$ 
final states\footnote{The 
study of Higgs production in association with a jet
was first suggested in the context 
 of improving
 $\tau$ reconstruction 
in the $\tau^+\tau^-$ decay channel \cite{tau}. 
}; 
this offers several advantages. 
The photons are more energetic than in the
case of the inclusive channel and 
the reconstruction of the jet in the calorimeter allows a
more precise determination of the interaction vertex, improving the efficiency and
mass resolution. Furthermore the existence of a jet in the final state allows
for a new type of  event selection and 
a  more efficient background suppression.
In addition, 
the necessary control of the background
contributions can probably be already achieved by the inclusion 
of the NLO corrections (the matrix elements
are known \cite{sigbern}).
In  a recent phenomenological
study \cite{adikss} it has been found that these advantages 
appear to be able to compensate 
the loss in production rates, provided one gets a
large $K$-factor also for this process.
The presentation of the NLO QCD corrections for this process is the main
purpose  of this letter.

The production process $gg\to H$ is given by loop diagrams 
in the Born approximation,  since the gluons interact  
with the Higgs boson via  virtual quark loops \cite{lo}. The exact calculation
of the NLO corrections is rather complex \cite{spira}.
Fortunately, the effective field theory approach \cite{approx} obtained in the
large top mass limit
with  effective gluon--gluon--Higgs coupling 
gives an accurate approximation (with or without QCD corrections) 
with an error less than 5\%,   
provided $m_{\rm H}\le 2\,  m_{\rm t}$ \cite{spira,five}.
It has been checked in LO,  by an explicit calculation,  that 
the approximation
remains valid also for the production of Higgs bosons with
large transverse momentum, provided  both $m_{\rm H}$ and $p_{\rm T}$ are
smaller than $m_{\rm t}$ \cite{glover}. 
It is therefore plausible to assume  that  the approximation remains valid 
also if we include NLO QCD corrections.
Recently, in this approximation and using the helicity method, 
 the transition amplitudes relevant to the NLO corrections have been
 analytically calculated for all the
contributing subprocesses
(loop corrections \cite{virt} and bremsstrahlung \cite{real}).  

The  available NLO  matrix elements
contain soft and collinear singularities
and therefore do not allow for a direct numerical evaluation of the
physical cross section.
In the past few years, exploiting the universal 
structure of the soft and collinear contributions, several efficient 
algorithms have been suggested  
to obtain 
finite cross section expressions from  the singular NLO matrix elements.
We have  used the   method of ref. \cite{fks} 
and implemented it into
 a numerical Monte Carlo style  
program which allows to calculate
 any infrared-safe physical quantity for
 the inclusive production of a Higgs boson with one jet    
in NLO accuracy.

In this paper we report some of our results obtained for
proton--proton collisions  with $\sqrt{S}=14\tev$. 
For the strong coupling constant at NLO (LO)
 we use the standard two-loop (one-loop)
 form with $\Lambda_{QCD}$ set to the value used in the analysis 
of the parton distribution function under consideration.
  Our default choice for the factorization and renormalization scales is 
$Q_0^2= (m_{\rm H}^2+p_{\rm T}^2)$, where $p_{\rm T}$ 
is  the transverse momentum of the Higgs boson. Here, unless stated, we
consider the case of $m_{\rm H}=120$ GeV and $p_{\rm T}>30$ GeV in the
kinematical region where the perturbative result can be applied without having to  consider low-$p_{\rm T}$ resummation effects.

Most of our curves  have been obtained 
with MRST (ft08a) parton distribution functions, but we
 will also show some results using CTEQ(4M) and GRV98 \cite{pdf}. 
To compare the leading with the NLO results, 
for consistency, 
 we  use the corresponding LO parton distributions from each set. 
We shall discuss  only results
 for the inclusive production cross section 
of a Higgs boson with large transverse momentum,
although as we mentioned above the Monte Carlo program allows
 to study   any infrared-safe quantity, 
including  the implementation of different jet algorithms 
and experimental cuts. 

\begin{figure}[htb]
\centerline{ \desepsf(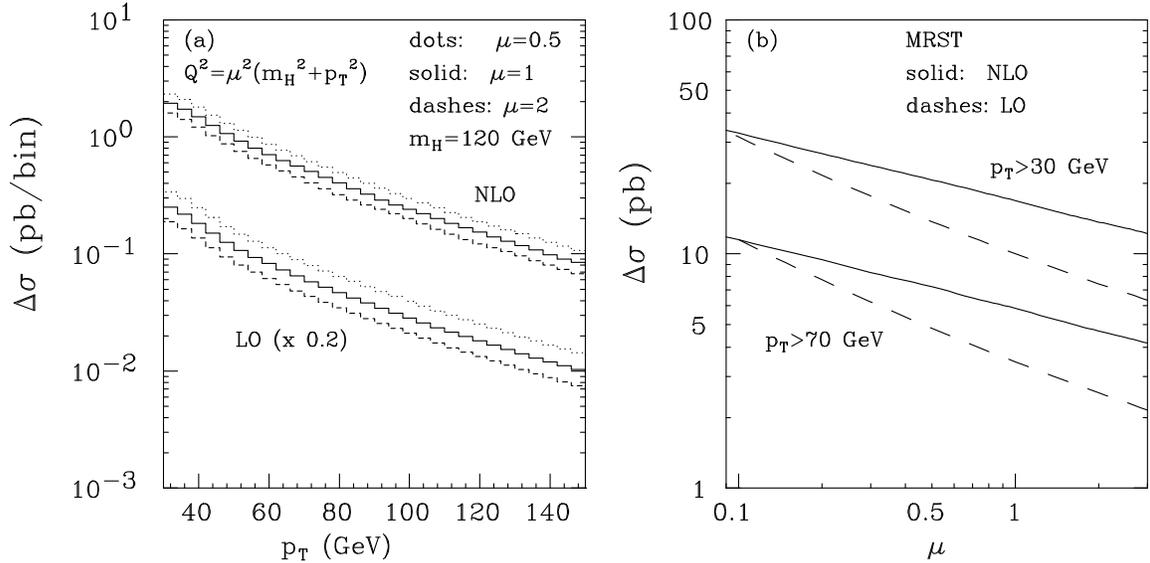 width 18 cm) } 
\caption{Scale dependence of NLO and LO
  distributions using MRST parton densities. (a) $p_{\rm T}$ distributions at
  different scales and (b) the scale dependence of the integrated cross sections  for $p_{\rm T}>30$ and  $70$ GeV} 
\end{figure}

In Fig. 1(a) we show the $p_{\rm T}$ distribution of 
the NLO and LO cross sections using MRST parton 
densities at three different scales $Q=\mu \, Q_0$, with $\mu=0.5,1,2$.
In this figure one can see three important points. First, 
the radiative corrections
are large; second, there is a 
 reduction in the scale dependence when going from LO to NLO; third,  the improvement
in the scale dependence
 is still not completely satisfactory.
The same features can be observed in more detail in Fig. 1(b),
 where the LO and NLO cross sections integrated 
for $p_{\rm T}$ larger than 30 and 70 GeV are shown 
as a function of the renormalization/factorization scale.
  Both the LO and NLO  cross sections
increase monotonically with decreasing   $\mu$ scale, 
down to the limiting value  where perturbative QCD can still 
 be applied. In NLO  the scale dependence has a maximum
and its position 
 characterizes the stability of the NLO perturbative  results.
 In our case, as a result of the very large
positive radiative corrections, 
 the position of the maximum is shifted down to small $\mu$ values
where the perturbative treatment is not
valid, indicating that the stability of the NLO result is not
completely satisfactory.
In the usual range of variation of $\mu$ from 0.5 to 2, however,  
the LO scale uncertainty amounts up to $\pm 35\%$, 
whereas at NLO this is reduced to $\pm 20\%$, 
indicating the relevance of the QCD corrections.
This feature  of the scale dependence in 
NLO is very much the same as the one in the case of 
inclusive Higgs production \cite{spira}.
\begin{figure}[htb]
\centerline{ \desepsf(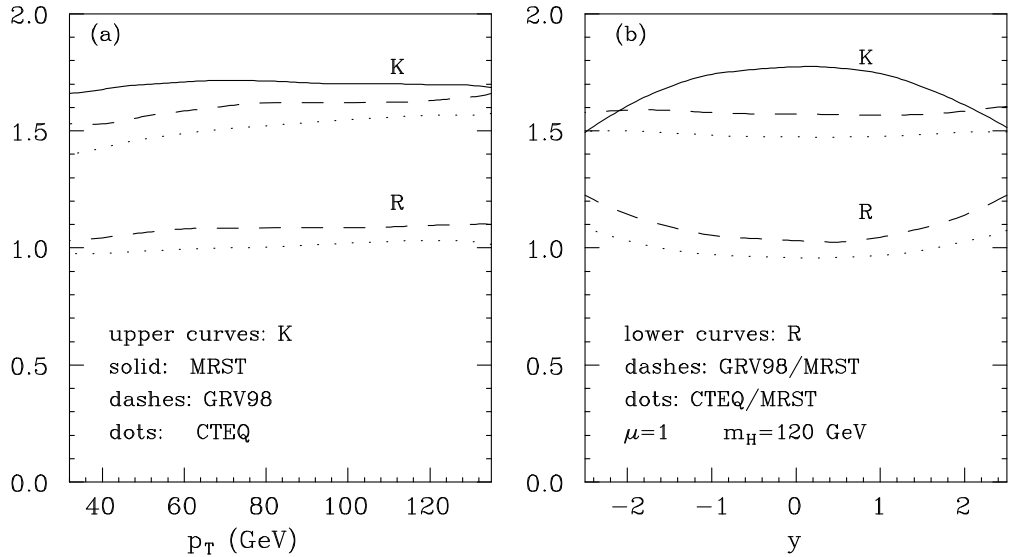 width 18 cm) } 
\caption{(a) $p_{\rm T}$  and (b) $y$ dependence of the $K$-factor and
the  ratio $R$ of the NLO cross sections 
for different sets of parton densities
(see eqs. (\ref{K}) and (\ref{R}))
.
 } 
\end{figure}
\newpage
In Fig. 2 we show the ratio
\begin{equation}
\label{K}
K=\frac{\Delta \sigma_{NLO}}{\Delta \sigma_{LO}}
\end{equation}
of the next-to leading and the LO cross sections ($K$-factor) for the three different sets of parton distributions: MRST, CTEQ and GRV98, as a function of the transverse momentum and the rapidity
of the Higgs boson.
 We can see that  the $K$-factor 
is  in the range 1.5--1.6 and it is  
almost constant (within 15\% accuracy)
 for a large range of $p_{\rm T}$ and $y$. 
In the $p_{\rm T}$ distribution 
the variation never exceeds 10\%, whereas 
it is  a bit larger for the $y$ distribution at large $|y|$.

The ratios of the NLO cross sections
\begin{equation}
\label{R}
 R=\frac{\Delta \sigma _{\rm CTEQ,GRV}}{\Delta \sigma _{\rm MRST}}
\end{equation}
computed by using CTEQ and GRV98 over the one obtained 
by using MRST parton densities are also shown. 
From there, it is possible to see that the differences in the $K$-factors basically
 come from variations  in the LO cross sections, mostly because of
 the value of
 $\Lambda_{QCD}$ used in each set.

We conclude that the properties of the
 $K$-factors found for large transverse momentum 
Higgs production are  very similar to
 the ones obtained for the total inclusive Higgs production.
They are about the same size, they show the same scale dependence, 
and  the $K$-factor changes mildly
with changing  $m_{\rm H},\, y$ and $p_{\rm T}$.
Its large value and its surprising independence from the kinematics
might be interpreted as an evidence for some universal origin of the large radiative corrections
 \cite{resum}. This requires further theoretical understanding.

\begin{figure}[htb]
\centerline{ \desepsf(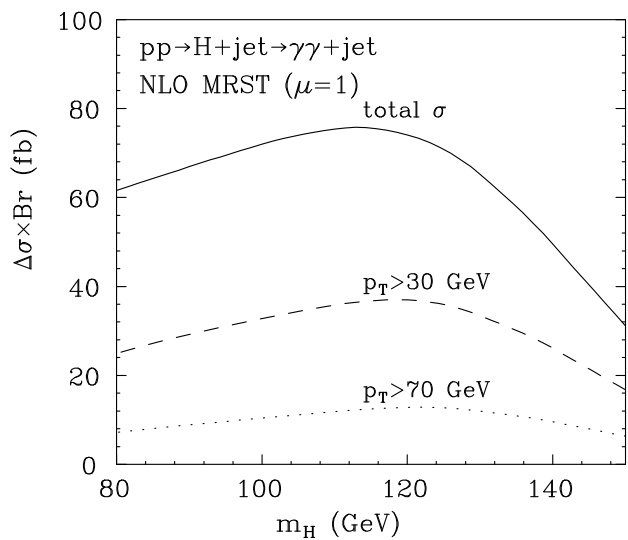 width 18 cm) } 
\caption{NLO-integrated cross section for the physics signal $p+p\to H + {\rm jet}\to
  \gamma+\gamma +$ jet for $p_{\rm T}> 30, 70$ GeV. We also include the value 
  for the total cross section ($p+p\to H\to \gamma+\gamma$).} 
\end{figure}

We have not included in our analysis the contributions from electroweak
reactions, which can increase the cross section  about 10\% when suitable cuts are applied \cite{adikss}. Nevertheless 
it is worth noticing that since QCD corrections to electroweak boson fusion are
substantially smaller than the ones corresponding to gluon fusion, the
significance of the electroweak contributions is
reduced at NLO.

In Fig. 3 we show  NLO
 cross section values for the physics signal
$p+p\to H+{\rm jet}\to \gamma+\gamma+{\rm jet}$  
as a function of the Higgs mass (with a reference value for the branching
ratio given by Br$(H\to \gamma
\gamma)= 2.18\cdot 10^{-3}$ for $m_{\rm H}=120$ GeV \cite{spiraprog}). 
For comparison,  
the cross section values of the physics
signal $p+p\to H\to \gamma+\gamma$ are also shown. From there it is possible to 
see that the loss in production rate due to the transverse momentum cut of
$p_{\rm T}>30$ GeV is less than a factor of 2 for the range of masses considered.

In conclusion, we have pointed out that, 
much as in the case of  inclusive Higgs production, the 
cross section values of the associated production of a Higgs boson with a jet
are increased by a $K$-factor of 1.5--1.6 given by NLO  
QCD radiative corrections. 
Our result confirms  previous suggestions 
that the production channel $p+p\to H+{\rm jet} $
 $\to \gamma+\gamma\, +$ jet
gives a measurable signal  for Higgs production at the LHC in the mass
range $100$--$140\gev$, 
crucial also for the ultimate test of the Minimal Supersymmetric
Standard Model. \\

We are grateful to M. Spira for discussions. One of us (DdeF) would like
to thank \\ S. Frixione for helpful comments.

\end{document}